\documentclass[12pt]{iopart}

\usepackage{graphicx}  
\begin{document}

\title[Ultracold Collisions of Polyatomic Molecules: CaOH]{Ultracold Collisions of Polyatomic Molecules: CaOH}

\author{Lucie D Augustovi\v{c}ov\'{a}$^{1}$ and John L Bohn$^{2}$ }

\address{$^{1}$Charles University, Faculty of Mathematics and Physics, Department of Chemical Physics and Optics, Ke Karlovu 3, CZ-12116 Prague 2, Czech Republic, \\
$^{1}$JILA, NIST, and Department of Physics, University of Colorado, Boulder, Colorado 80309-0440, USA.}

\ead{augustovicova@karlov.mff.cuni.cz}
\vspace{10pt}

\begin{abstract}
Ultracold collisions of the polyatomic species CaOH are considered, in internal states where the collisions should be dominated by long-range dipole-dipole interactions.  The computed rate constants suggest that evaporative cooling can be quite efficient for these species, provided they start at temperatures achievable by laser cooling.  The rate constants are shown to become more favorable for evaporative cooling as the electric field increases.  Moreover, long-range dimer states (CaOH)$^*_2$ are predicated to occur, having lifetimes on the order of microseconds.
\end{abstract}

%
\noindent{\it Keywords}: ultracold molecules, evaporative cooling, Stark effect, dipole-dipole interaction
%
%
\maketitle
%
%

\section{Introduction}

The technology to laser cool molecules leads the way to a wave of truly ultracold molecular species, achieving temperatures on the microKelvin scale rather than the milliKelvin scale \cite{McCarron18_JPB}.  These temperatures are low enough that the molecules can be confined in magnetic  \cite{Williams18_PRL,McCarron18_PRL} or optical dipole traps \cite{Anderegg18_NP,Cheuk18_PRL}, can be produced in individual quantum states, tend to collide in individual partial waves, and have collisions that respond strongly  to laboratory electric and magnetic fields \cite{Krems_book}. These are all ingredients that enhance the experimental ability to control ultracold molecules.  The newest members on the list of laser-coolable species are polyatomic species \cite{Kozyryev16_CPC,Kozyryev17_PRL1}.  The linear triatomic species SrOH is a good candidate for laser cooling \cite{Li19_preprint}, and has been deflected by optical forces \cite{Kozyryev18_PRL}, opening the way for similar species such as CaOH and CaOCH$_3$ \cite{Kozyryev19_NJP}, and more besides.  These species should expand opportunities for quantum information, sensing, and fundamental physics \cite{Kozyryev16_CPC}.

Central to the properties of an ultracold gas are the collision cross sections of its constituent molecules.  As in any ultracold environment, high elastic scattering rates are desirable to bring the gas to thermal equilibrium, while low inelastic scattering  rates are essential to protect the gas from two-body losses.  Understanding collision cross sections and their response to applied electromagnetic fields is also vital for controlling collisions, with attendant applications to ultracold chemistry.  The species SrOH has been studied experimentally in collisions with helium buffer gas atoms at 2.2 K, finding that vibrational quenching occurs rapidly in these collisions \cite{Kozyryev15_NJP}.  In addition, collisions of SrOH with lithium atoms has been investigated theoretically, concluding that sympathetic cooling of the molecule with this atom is feasible \cite{Morita17_PRA}.

In this article we extend cold collision theory of linear polyatomic molecules, considering CaOH molecules colliding with each other.  Central to our approach is that, for certain collisions at ultralow temperature, the scattering rates and their field dependence rely on physics that occurs when the molecules are far apart, that is, on scales larger than the range of the exchange potentials between them. This circumstance simplifies the description of scattering, and leads to certain common behaviors.   In this article, exploiting the electric dipole moment of CaOH and considering a state that has a small parity doublet, we find that these  behaviors still occur.  The ones that we single out are: 1) a suppression of inelastic scattering at sufficiently high electric field and sufficiently low temperature, for states that can be optically trapped; and 2) a set of electric-field resonances, previously described as ``field linked states,'' \cite{Avdeenkov03_PRL,Avdeenkov04_PRA} that could serve as an additional platform for controlling these species and their interaction.

\section{The Molecule}

The molecule CaOH (or the closely related SrOH) has a linear geometry in its $^2 \Sigma^+$ electronic ground state.  Around this linear geometry, the molecule has vibrational modes in the Ca-O and O-H bonds, denoted by quantum numbers $\nu_1$ and $\nu_3$, respectively; and a bending vibration denoted $\nu_2$.  The collective state of the vibration is then labeled $(\nu_1,\nu_2,\nu_3)$ \cite{Townes}.  The bending mode is lower in energy in these molecules, whereby  at low temperatures we focus on the states $(0,\nu_2, 0)$.  

For small vibrational quanta $\nu_2$, we regard the molecule as a rigid asymmetric rotor, defined by a principal axis that we think of as the Ca-H axis.  Owing to the bending vibration, the O atom is  displaced a small distance off this axis.  Suitable linear combinations of vibrations in the molecule-fixed $x$ and $y$ directions amount to rotation of the O atom around the molecular axis, with component $l$ on this axis, where $l$ is a signed integer.  If the electron were to have angular momentum projection $\Lambda$ on this axis, then the relevant quantum number in Hund's case a would be $K = l + \Lambda$, but for the $\Sigma$ electronic state, $\Lambda=0$ and $K=l$.  To specify the value of $|l|$ given $\nu_2$, one writes the vibrational state as $(\nu_1,\nu_2^{|l|},\nu_3)$.

The asymmetric rigid  rotor is therefore described using the usual rotor wave functions
\begin{eqnarray}
\langle \alpha \beta \gamma | l N  M_N \rangle = 
 \sqrt{ \frac{ 2N+1 }{ 8\pi^2 } } D^{N*}_{M_N l}( \alpha \beta \gamma )
\end{eqnarray}
in terms of the Euler angles $(\alpha,\beta,\gamma)$ giving the orientation of the molecule.  The vibrational rotation quantum  number $l$, tied to the molecular axis,  is treated like one would treat the projection of the electron angular momentum in Hund's case a.  This extends even to notation: states with angular momenta $|l|=0,1,2...$ are labeled $^2\Sigma$, $^2\Pi$, $^2\Delta$ ... (It is understood that the electron remains in a $\Sigma$ state.)  For $|l|>0$, the degeneracy of two states is broken, producing an $l$-doubling analogous to $\Lambda$-doubling in a case a molecule.   Meanwhile, the electronic and nuclear spin states are well described by Hund's case b.  An appropriate uncoupled basis state for the rotor wave functions is then
\begin{eqnarray}
|l NM_N \rangle |SM_S \rangle |I M_I \rangle.
\label{eq:uncoupled_basis}
\end{eqnarray}
This basis forms the foundation upon which all the results are computed and interpreted in what follows.  Appropriate basis sets may, however, require different superpositions of these states in the low- and high-electric field limits.

In this paper we focus on a particular state, the lowest bending excitation with $\nu_2=1$.  This is because it is the lowest-lying state with an $l$-doublet, and hence can be polarized easily in a small electric field.  Thus the low temperature scattering behavior is expected to be dominated by dipolar forces between the molecules,  enabling control over the collisions.  

\subsection{Field-Free Hamiltonian}

In the absence of an applied field, the states (\ref{eq:uncoupled_basis}) are coupled into a total angular momentum scheme, first adding ${\bf N}$ and ${\bf S}$ to produce ${\bf J}$, then adding ${\bf I}$ to produce the total spin ${\bf F}$:
\begin{eqnarray}
\fl
|l [(NS)JI]FM_F \rangle = \!\!\!\sum_{M_NM_SM_I}\!\!\! |l NM_N \rangle |SM_S \rangle |I M_I \rangle
\langle NM_N SM_S | JM_J \rangle \langle JM_J IM_I | FM_F \rangle.
\end{eqnarray}
When $l \ne 0$, these states are combined into parity eigenstates
\begin{eqnarray}
\fl
| |l| , \epsilon[(NS)JI]FM_F  \rangle = \frac{ 1 }{ \sqrt{2} } \Big[ |(l) [(NS)JI]FM_F \rangle + \epsilon |(-l) [(NS)JI]FM_F \rangle \Big]
\end{eqnarray}
 with parity $p=\epsilon(-1)^{N-l}$.  The complete basis set is then
\begin{eqnarray}
|(\nu_1,\nu_2^{|l|},\nu_3); |l|,\epsilon; [(NS)JI]FM_F  \rangle.
\label{eq:parity_basis}
\end{eqnarray}

For a given electronic state and a given vibrational state $(\nu_1,\nu_2^{|l|},\nu_3)$, the Hamiltonian of the molecule is written as a sum of several terms, in roughly descending order of energy:
\begin{eqnarray}
H = H_{\rm vib} + H_{\rm rot} + H_{\rm sr} + H_{\rm ld} + H_{\rm hf},
\end{eqnarray}
denoting, respectively, the vibrational and rotational energies, the spin-rotation coupling, the $l$-doubling, and the hyperfine interaction.  

For $\nu_2=1$ states, the model Hamiltonian $H$ is diagonal in the basis chosen, with  the  matrix elements as given in \cite{Fletcher95_JCP}.  In higher states this is not the case, for example, for $\nu_2=2$ there can be mixing between the $^2\Sigma$ and $^2\Delta$ states, but this will not concern us here.  The rotational Hamiltonian, ignoring centrifugal distortion,  is
\begin{eqnarray}
H_{\rm rot} = B_v [ N(N+1)-l^2].
\end{eqnarray}
The spin-rotation Hamiltonian, again ignoring centrifugal distortion, is diagonal in the basis (\ref{eq:parity_basis}) and is given by
\begin{eqnarray}
H_{\rm sr} = \gamma {\bf N} \cdot {\bf S} = \frac{ \gamma }{ 2 } [ J(J+1) - N(N+1) - S(S+1) ].
\end{eqnarray} 
 The $l$-doubling arises due to Coriolis coupling of the state $l$ to states with $l \pm 1$, and grows with $N$.  Using the conventions established in Refs.~\cite{Fletcher95_JCP,Kopp67_CJP}, the $l$-doubling Hamiltonian is diagonal in (\ref{eq:parity_basis}), with matrix elements
  \begin{eqnarray}
 H_{\rm ld} = \frac{ q _l\epsilon }{ 2 } N(N+1).
 \end{eqnarray}
 These states are labeled by the letters $e$ and $f$, assigned by the convention
  \begin{eqnarray}
 \label{parity_states} 
 p =  \left\{ \begin{array}{ll} +(-1)^{J-1/2}, \;\;\; e \\ -(-1)^{J-1/2}, \;\;\; f \end{array} \right.
 \end{eqnarray}
  Finally, the hyperfine interaction is the smallest perturbation to the molecule, taking the form
  \begin{eqnarray}
H_{\rm hf} = b {\bf J} \cdot {\bf I} = \frac{ b }{ 2 } [ F(F+1) - J(J+1) - I(I+1) ],
\end{eqnarray}
where $I=1/2$ is the spin of the hydrogen atom, the only relevant nuclear spin in the $^{40}$Ca$^{16}$OH molecule.

Most of the spectroscopic constants are reported in Ref.~\cite{Fletcher95_JCP}.  For the $|l|=1$ state of CaOH, we use $B=9996.75184$ MHz, $\gamma =35.051$ MHz, $q_l=-21.6492$ MHz.  The hyperfine constant has not been measured, to our knowledge.  We therefore use the value  measured for the $l=0$, $N=1$ level, $E_{F=1}-E_{F=0} = 7 \times 10^{-3}$MHz  \cite{Scurlock93_JMS}.  

\subsection{The Electric Field}

Polar molecules like CaOH will obviously respond to an electric field.  A magnetic field is perhaps less relevant at this point, since the electron is only weakly coupled to the molecular axis.  We therefore focus on the Stark effect.  Its Hamiltonian is
\begin{eqnarray}
H_{\rm E} = - {\bf d} \cdot {\cal E} = -d{\cal E} C_{10}(\cos \beta),
\end{eqnarray}
where $\beta$ is the angle between ${\bf d}$ (which coincides with the molecular axis) and ${\cal E}$.  For use later on, we compute the matrix elements of $C_{1q}$ for arbitrary $q$.  These are given in terms of the reduced matrix element as
\begin{eqnarray}
\label{eq:mol_matrix_element}
&& \langle l,\epsilon; [(NS)JI]FM_F | C_{1q} | l,\epsilon^{\prime}; [(N^{\prime}S)J^{\prime}I]F^{\prime}M_F \rangle \nonumber \\
&& \;\;\;\;\; = (-1)^{F-M_F} \sqrt{2F+1} 
\left( \begin{array}{ccc} F & 1 & F^{\prime} \\ -M_F & q & M_F^{\prime} \end{array} \right)  \\
&& \;\;\;\;\;\;\;\; \times 
 \langle l,\epsilon; [(NS)JI]F || C_{1} || l,\epsilon^{\prime}; [(N^{\prime}S)J^{\prime}I]F^{\prime} \rangle.
 \nonumber
\end{eqnarray}
The reduced matrix element is computed in the usual way \cite{BS,Ticknor05_PRA}
\begin{eqnarray} 
&& \langle l,\epsilon; [(NS)JI]F || C_1 || l,\epsilon^{\prime}; [(N^{\prime}S)J^{\prime}I]F^{\prime} \rangle \nonumber \\
&& \;\;\;\;\; = 
 (-1)^{F^{\prime}+J+J^{\prime}+S+I+l}
\left( \frac{ 1 + \epsilon \epsilon^{\prime}(-1)^{N+N^{\prime}+1} }{  2 } \right)  [F^{\prime}][J][J^{\prime}][N][N^{\prime}]\nonumber \\
&& \;\;\;\;\;\;\;\; \times
\left\{ \begin{array}{ccc} J & J^{\prime} & 1 \\ N^{\prime} & N & S \end{array} \right\}
\left\{ \begin{array}{ccc} F & F^{\prime} & 1 \\ J^{\prime} & J & I \end{array} \right\}
 \left( \begin{array}{ccc} N & 1 & N^{\prime} \\ -l & 0 & l \end{array} \right),
\end{eqnarray}
where $[J] = \sqrt{2J+1}$, etc.  The magnitude of the dipole moment has been measured as $d=1.465$ D \cite{Steimle92_JCP}.  

\subsection{High-Field Limit}
\label{HF}

Although the scattering calculations presented below are performed by casting the two-body Hamiltonian in the zero-field basis (\ref{eq:parity_basis}), to describe the states in the high-field limit it is useful to specify the quantum numbers that are good there.  In this limit the dominant term in the Hamiltonian is the Stark effect, which  in the uncoupled basis (\ref{eq:uncoupled_basis}) has matrix elements diagonal in $N_M$ and $l$, as well as the spins:
\begin{eqnarray}
&& \langle l N M_N | \langle S M_S | \langle I M_I | -d \cdot {\cal E} | 
l^{\prime} N^{\prime} M_N^{\prime} \rangle | S M_S^{\prime} \rangle |I M_I^{\prime} \rangle \\
&& = -d {\cal E} (-1)^{M_N-l} [ N ][ N^{\prime} ]
\left( \begin{array}{ccc} N & 1 & N^{\prime} \\ -M_N & 0 & M_N \end{array} \right)
\left( \begin{array}{ccc} N & 1 & N^{\prime} \\ -l & 0 & l \end{array} \right)
\delta_{M_SM_S^{\prime}} \delta_{M_IM_I^{\prime}}. \nonumber 
\end{eqnarray}
Moreover, we are primarily interested in the $N=1$ rotational ground state.  This state is mixed with the nearby $N=2$ state for fields on the order of $4B_v/d \approx 5 \times 10^{4}$ V/cm.  So long as we remain well below this field, the Stark Hamiltonian is  diagonal in $N$ as well, and its matrix elements simplify to
\begin{eqnarray}
&& \langle l N M_N | \langle S M_S | \langle I M_I | -d \cdot {\cal E} | 
l^{\prime} N M_N^{\prime} \rangle | S M_S^{\prime} \rangle |I M_I^{\prime} \rangle \nonumber \\
&& = -d {\cal E} \frac{ l M_N }{ N(N+1) } \delta_{M_SM_S^{\prime}} \delta_{M_IM_I^{\prime}},
\label{eq:Stark_diagonal}
\end{eqnarray}
which harbors a double degeneracy for each value of $lM_N$.  

Next is the spin-rotation Hamiltonian, which couples different $M_N$ and $M_S$.  Since
\begin{eqnarray}
H_{\rm sr} = \gamma {\bf N} \cdot {\bf S} = \sum_q (-1)^q N_{q}S_{-q},
\end{eqnarray}
we have in the uncoupled basis
\begin{eqnarray}
\label{eq:spin-rotH}
&& \langle l N M_N | \langle S M_S | \langle I M_I |H_{\rm sr} | 
l^{\prime} N M_N^{\prime} \rangle | S M_S^{\prime} \rangle |I M_I^{\prime} \rangle \nonumber \\
&& = \gamma (-1)^{q+N-M_N+S-M_S} \sqrt{N(N+1)(2N+1)S(S+1)(2S+1) } \\
&& \quad \times \left( \begin{array}{ccc} N & 1 & N \\ -M_N & q & M_N^{\prime} \end{array} \right)
  \left( \begin{array}{ccc} S & 1 & S \\ -M_S & -q & M_S^{\prime} \end{array} \right)
  \delta_{M_IM_I^{\prime}}, \nonumber
\end{eqnarray}
with $q = M_N - M_N^{\prime} = M_S^{\prime}-M_S$.  
Finally, the $l$-doubling Hamiltonian is off-diagonal in the $l$ quantum number:
\begin{eqnarray}
&& \langle l N M_N | \langle S M_S | \langle I M_I |H_{\rm ld} | 
l^{\prime} N M_N^{\prime} \rangle | S M_S^{\prime} \rangle |I M_I^{\prime} \rangle \nonumber \\
&& = \left( 1-\delta_{ll^{\prime}} \right) q_l N(N+1) \delta_{M_NM_N^{\prime}}
\delta_{M_SM_S^{\prime}} \delta_{M_IM_I^{\prime}}
\end{eqnarray}
The hyperfine Hamiltonian is even smaller, and we will not call it out in the high-field limit.  

The Stark effect of the $(0,1^1,0)$, $N=1$ levels is shown in Figure 1.  The range shown spans the transition from low- to high-field behavior, which occurs to due  mixing of the zero-field parity states by the electric field.  The transition between these two limits occurs at a field of approximately ${\cal E}_0 = 2|q_l|/d \approx$ $58$ V/cm. 
States in the zero-field limit, ${\cal E} < {\cal E}_0$, are labeled by their good quantum numbers $J$, $F$, and parity.  States in the high-field limit, ${\cal E} > {\cal E}_0$, split into three main groups, characterized by the value of $l M_N$, that rise in energy with field, fall with energy, or remaining relatively constant, in accordance with (\ref{eq:Stark_diagonal}).  These states are further split by $H_{\rm sr}$ and $H_{\rm ld}$.  Diagonal elements of $H_{\rm sr}$ allow the identification of the dominant values of $M_N$, whereby these states can be labeled  by the value of $M_J$.  Also shown is the total projection of angular momentum, $M_F$, which will allow us to identify spin-stretched states where necessary. 

For simplicity of notation, the  six relevant manifolds shown at high field are labeled simply in order of increasing energy as $a$, $b$, $c$, $d$, $e$, $f$. Here $e$ and $f$ do not have the usual parity meaning (Eq. (\ref{parity_states})), but are merely putting energies in order, as shown. In what follows, it will be relevant to describe scattering events in terms of the rotation and spin quantum numbers, along with the fine structure manifold.  Thus we will employ the shorthand notation for the used basis set
\begin{eqnarray}
|x, l; M_N M_S \rangle,
\label{eq:shorthand}
\end{eqnarray}
where $x=a, b, \dots f$.  If needed, we will also specify the total spin $M_F$, but the nuclear spin plays a minor role in scattering.

\begin{figure}[h]
\center
\includegraphics[width=0.8\textwidth]{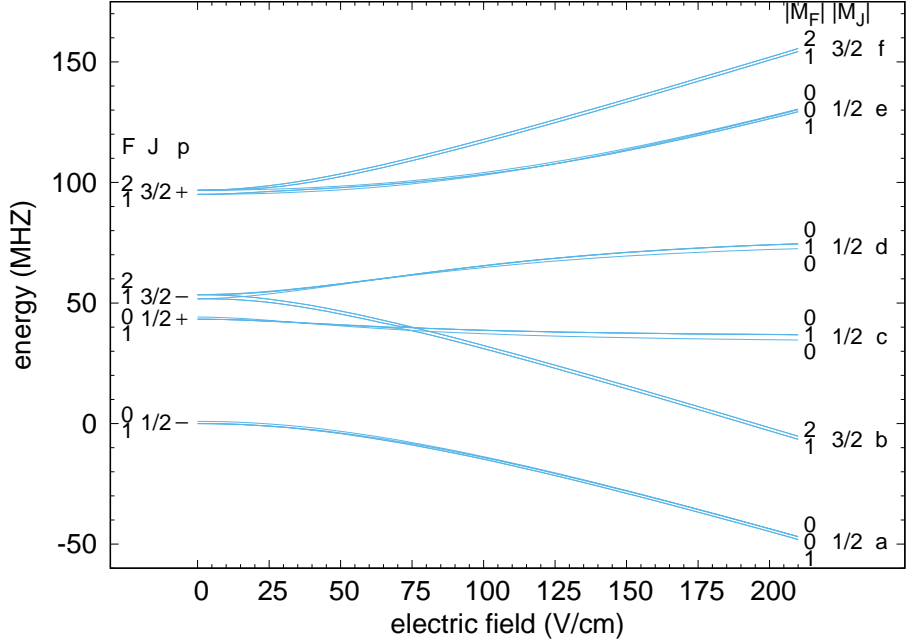}
\caption{Stark effect in the $(0,1^{|l|=1},0)$, $N=1$ state of CaOH.  At zero field, the states are labeled by the total electron-plus-rotation angular momentum $J$, the total spin $F$, and the parity $p$;  at larger electric field the states are labeled by the projections $M_J$ and $M_F$ of these angular momenta along the field axis.  Each line is doubly degenerate in $lM_F$.  As a shorthand, the fine structure manifold at high field are labeled by the indices $a-f$.}
\end{figure}


\section{The Scattering Hamiltonian}

At ultralow collision energies, we focus on the long-range interactions between the molecules.  This includes a van der Waals interaction $-C_6/R^6$, which we take to be isotropic. Scattering at long range is driven by  the dipole-dipole interaction,
\begin{eqnarray}
V_{\rm d} = - \frac{ \sqrt{30} d^2 }{ R^3 } \sum_{qq_1q_2}
\left( \begin{array}{ccc} 2 & 1 & 1 \\ q & -q_1 & -q_2 \end{array} \right)
C_{2-q}(\theta \phi) C_{1q_1}(\beta_1 \alpha_1) C_{1q_2}(\beta_2 \alpha_2).
\end{eqnarray}
Here $(\theta, \phi)$ are the polar angles of the intermolecular vector ${\bf R}$, and $(\beta_i,\alpha_i)$ are the polar angles giving the orientation of molecule $i$.  

For a pair of molecules, the unsymmmetrized, low-field  basis functions are written
\begin{eqnarray}
\fl
|\eta_1 F_1 M_{F_1} \rangle |\eta_2 F_2 M_{F_2} \rangle |LM_L \rangle
\equiv | l,\epsilon_1 [(N_1S)J_1I]F_1 M_{F_1} \rangle | l,\epsilon_2 [(N_2S)J_2I]F_2 M_{F_2} \rangle
|L M_L \rangle.
\end{eqnarray}
The matrix elements of the interaction are then given by
\begin{eqnarray}
&& \langle \eta_1 F_1 M_{F_1} | \langle \eta_2 F_2 M_{F_2} | \langle L M_L | V_{\rm d} 
|\eta_1^{\prime} F_1^{\prime} M_{F_1}^{\prime} \rangle | \eta_2^{\prime} F_2^{\prime}M_{F_2}^{\prime} \rangle
|L^{\prime} M_L^{\prime} \rangle \nonumber \\
&& \;\;\;\;\;\;  = - \frac{ \sqrt{30} d^2 }{ R^3 } 
\left( \begin{array}{ccc} 2 & 1 & 1 \\ q & -q_1 & -q_2 \end{array} \right) 
\langle LM_L | C_{2-q} | L^{\prime} M_L^{\prime} \rangle \\
&& \;\;\;\;\;\;\;\;\;\; \times 
 \langle \eta_1 F_1 M_{F_1} | C_{1q_1} | \eta_1^{\prime} F_1^{\prime} M_{F_1}^{\prime} \rangle 
  \langle \eta_2 F_2 M_{F_2} | C_{1q_2} | \eta_2^{\prime} F_2^{\prime} M_{F_2}^{\prime} \rangle.
  \nonumber
\end{eqnarray}
The matrix elements within molecular states are given by (\ref{eq:mol_matrix_element}), while the partial wave matrix element is
\begin{eqnarray}
\langle LM_L | C_{2-q} | L^{\prime} M_L^{\prime} \rangle = (-1)^{M_L} [L][L^{\prime}]\!
\left( \begin{array}{ccc} L & 2 & L^{\prime} \\ 0 & 0 & 0 \end{array} \right)\!\!
\left( \begin{array}{ccc} L & 2 & L^{\prime} \\ -M_L & -q & M_L^{\prime} \end{array} \right)
\label{eq:L_mat_elt}
\end{eqnarray}
In practice, matrix elements in this basis are transformed into the basis of eigenstates in the desired electric field, and form the physical scattering states.  The matrix elements must moreover be symmetrized for particle exchange.  In what follows here, we will consider  molecules colliding in initially identical quantum states, whereby we consider only even partial waves for these bosonic molecules.
 
 Note also there is a possibility that these molecules react chemically.  In particular, the reaction
 \begin{eqnarray}
 {\rm CaOH} + {\rm CaOH} \rightarrow {\rm Ca(OH)}_2 + {\rm Ca}
 \end{eqnarray}
is exothermic by some 13,000 K.  Ca(OH)$_2$ is a stable compound used in industrial applications like paper production and sewage treatment.  It is not known whether this reaction occurs at low temperatures in the gas phase.  If it does, this is obviously a detriment to producing and maintaining a stable, ultracold gas of CaOH.  For these reasons we will disregard the possibility of the reaction, as the potential energy surface is unknown, and focus instead on ultracold collisions where the molecules are expected to be shielded by the repulsive parts of the dipole-dipole interaction.  

\subsection{Scattering Calculations} 

Calculations of collision cross sections are performed by first casting the Hamiltonian into the low-field basis as described above.  The Hamiltonian of the two molecules is diagonalized in the presence of the applied field, if any, to define the asymptotic scattering channels.  The incident channel selects one of these to describe the states of the colliding molecules.  The molecules are identical bosons, so if we consider scattering two molecules in identical initial states, we incorporate even partial waves $L=0,  \dots, L_{\rm max}$.  We find the calculations are converged with $L_{\rm max}=18$.  

In practice, inelastic collisions of polar molecules are subject to propensity rules that favor small values of $\Delta M_L$, that is, it is difficult to change the projection of orbital angular momentum significantly.  This propensity was explored in Ref.~\cite{Augustovicova18_PRA}.  In  the calculations that follow, we restrict that basis set to $|\Delta M_L| \le 3$.  This results in a set of typically $\sim 10^3$ channels per scattering calculation.  We perform this calculation using a log-derivate propagator method \cite{Johnson}.  The total cross section is a sum of partial cross sections over all even incoming partial-wave angular momentum, $\sigma_{i\rightarrow f}(E) = \sum_L \sigma_{L,i\rightarrow f}(E)$.
 
 \section{Results}
 
We here report two significant properties of ultracold collisions of $(0,1^1,0)$ CaOH molecules, at least among those that are dominated by long-range physics.  The first is the possibility of evaporative cooling in an appropriate state.  The second is the occurrence of field-linked states, short-lived dimers consisting of a pair of CaOH molecules weakly bound by dipolar forces.  
 
  \subsection{Prospects  for Evaporative Cooling in the b State}
  
  Evaporative cooling is efficient only to the extent that elastic collisions occur at far higher rates than inelastic collisions. One good strategy for reducing inelastic collision rates is to never let the molecules get close together. This idea is illustrated in Figure 2, which shows a simplified version of the adiabatic potential energy curves between the molecules, for molecules initially in the $a$ or $b$ fine structure manifolds (in the notation of Figure 1), and in a field of ${\cal E} = 6000$ V/cm.  
  
  Molecules in the $a$ manifold would have no lower-energy fine structure state to scatter into, and thus are immune to fine-structure-changing collisions.  However,  the lowest $L=0$ partial wave adiabatic curve is attractive and encourages the molecules to ``go into the lion's den'' at small $R$, where they may react chemically or else suffer the vibrational transition $\nu_2 =1 \rightarrow \nu_2=0$. 
 
\begin{figure}[h!]
 \center
\includegraphics[width=0.8\textwidth]{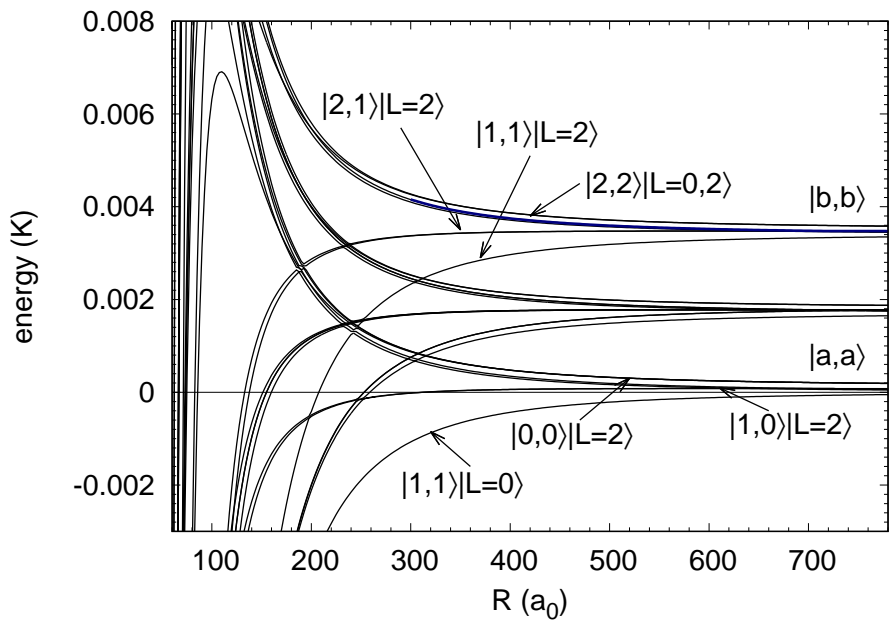}
\caption{Selected adiabatic potential energy curves for long-range CaOH-CaOH potentials at 6\,000 V/cm electric field.  These curves are simplified for clarity by including only the partial waves $L=0,2$ in their construction, and include only those curves correlating to the fine structure manifolds $a$ and $b$ at long range.  Each channel is labeled by the total spin projection quantum numbers, along with the partial wave component, $|M_{F1},M_{F2}\rangle|LM_L \rangle$.  The incident channel $|2,2\rangle|L=0,M_L=0\rangle$, with molecules in the spin-stretched state and correlating to the $|bb\rangle$ fine structure threshold, is highlighted. This is the incident channel for the rate coefficients presented in Figure 3.} 
\label{AdiabC-ab}
\end{figure}

 The situation is different for molecules in the fine structure states $b$.  Adiabatic curves for the this limit, in the spin-stretched states $|M_{F_1}M_{F_2}\rangle = |22 \rangle$ are repulsive, as seen in Figure \ref{AdiabC-ab}.  This repulsion originates in the dipole-dipole interaction inducing couplings to the lower energy states.  Level repulsion ensures that the upper states rise in energy at smaller $R$ where the dipole-dipole interaction grows in strength.  This is the principle of electrostatic shielding \cite{Avdeenkov06_PRA,Wang15_NJP,Quemener16_PRA,Gonzalez17_PRA}.

 \begin{figure}[h!]
 \center
\includegraphics[width=0.8\textwidth]{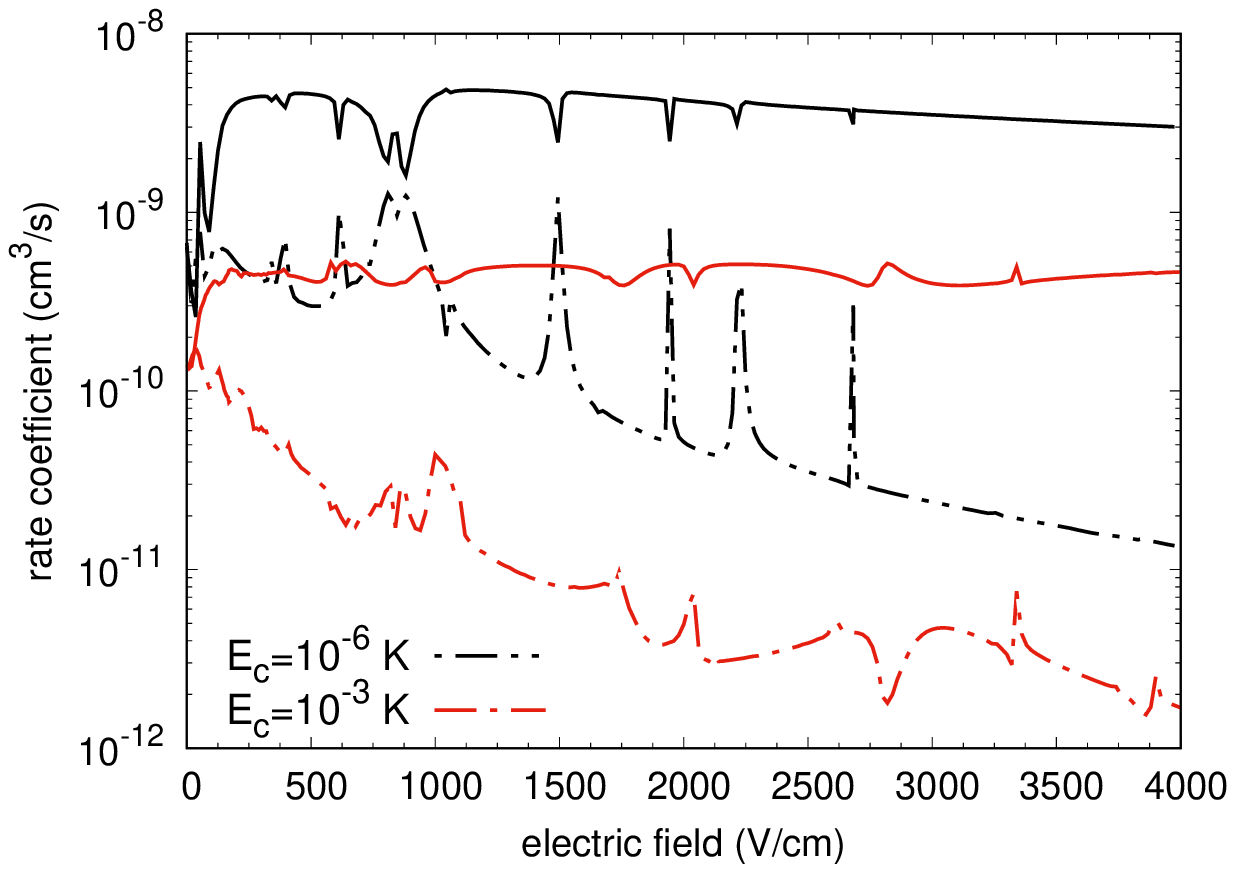}
\caption{Rate coefficients for elastic (solid curves) and inelastic (dashed curves) scattering as a function of electric field. The collision is initiated in the states $|b,l=1;M_F=2\rangle$ of molecules at two different collision energies $E_c = 1 \mu$K (black lines) and $E_c = 1$ mK (red lines).}
\label{rateE10-36}
\end{figure}

 We therefore focus on  spin-stretched molecules with $M_F=2$ in the $b$ state.  Figure \ref{rateE10-36} shows  rate coefficients versus field strength at two different collision energies, 1 $\mu$K and 1 mK.  The elastic rate constants remain high at all values of electric field, due to generically strong scattering of dipoles.  A remarkable feature  is an overall decreasing trend of loss rates with applied electric field at both energies.  As a rule of thumb, evaporation is efficient when $K_{\rm el}/K_{\rm inel} \ge 100$, which occurs for experimentally reasonable fields. Our calculations indicate that for $E_c = 1$ mK such a field is $\mathcal{E}\sim 3\,500$ V/cm and for $E_c = 1 \mu$K it is $\mathcal{E}\sim 2\,500$ V/cm.

The cause of this suppression of $K_{\rm inel}$ at high electric fields is described in detail in Refs.~\cite{Augustovicova18_PRA,Augustovicova17_PRA}, which estimates transition amplitudes  in the Born approximation.  
Central to this approximation is the proportionality 
\begin{eqnarray}
T_{\rm initial, final} \propto \langle {\rm initial} | C_3 |{\rm final} \rangle,
\end{eqnarray}
where $T$ is the transition matrix element between initial and final scattering channels, and $\langle{\rm initial} | C_3 |{\rm final} \rangle$ is the matrix element of the dipole coupling between the field-dressed initial and final states.  Not shown explicitly here is a radial integral over the scattering wave functions.  Selection rules for the direct transitions in the first Born approximation reside in the angular factor $C_3$.

\begin{table}[h!]
\caption{\label{ab_qn}Selected quantum numbers of states in the lowest two fine structure manifolds.  For the scattering calculations, the state in the first line is the incident channel.} 

\begin{indented}
\lineup
\item[]\begin{tabular}{c@{\extracolsep{11pt}}*{3}{c}}
\br  
manifold & $l$ & $M_N$ & $M_S$ \\                            
\mr
$b$ & 1 & 1 & 1/2 \cr
$b$ & $\-1$ & $\-1$ & $\-1/2$ \cr
\hline
$a$ & 1 & 1 & $\-1/2$ \cr
$a$ & $\-1$ & $\-1$ & 1/2 \cr
\br
\end{tabular}
\end{indented}
\end{table}

Quantum  numbers for states in  the $a$ and $b$ fine structure manifolds are given in Table \ref{ab_qn}; we disregard the nuclear spin as a spectator degree of freedom and denote the states as in (\ref{eq:shorthand}).  In this uncoupled basis relevant at high electric field, angular matrix elements of the dipole-dipole interaction read
\begin{eqnarray}
\fl
 \langle l_1 N_1 M_{N_1} S M_{S_1} | \langle l_2 N_2 M_{N_2} S M_{S_2} |\langle LM_L |C_3 
|l_1^{\prime} N_1^{\prime} M_{N_1}^{\prime} S M_{S_2}^{\prime}\rangle
|l_2^{\prime} N_{2}^{\prime} M_{N_2}^{\prime} S M_{S_2}^{\prime} \rangle |L^{\prime} M_L^{\prime} \rangle
\nonumber \\
= 
\left( \begin{array}{ccc} 2 & 1 & 1 \\ q & -q_1 & -q_2 \end{array} \right)
\langle LM_L | C_q | L^{\prime} M_L^{\prime} \rangle \\
 \;\;\;\;\;\;\;\; \times \langle  l_1 N_1 M_{N_1} | C_{q_1} | l_1^{\prime} N_1^{\prime} M_{N_1}^{\prime} \rangle
\langle  l_2 N_2 M_{N_2} | C_{q_2} | l_2^{\prime} N_2^{\prime} M_{N_2}^{\prime} \rangle
\delta_{M_{S_1}M_{S_1}^{\prime}}  \delta_{M_{S_2}M_{S_2}^{\prime}}, \nonumber
\end{eqnarray}
where $q=M_L - M_L^{\prime}$, $q_1 = M_{N_1}^{\prime} - M_{N_1}$, $q_2 = M_{N_2}^{\prime} - M_{N_2}$; the matrix element in partial wave quantum numbers is given in (\ref{eq:L_mat_elt}); and the molecular matrix elements are given by
\begin{eqnarray}
\fl
\langle  l_i N_i M_{N_i} | C_{q_i} | l_i^{\prime} N_i^{\prime} M_{N_i}^{\prime} \rangle
&=&  (-1)^{M_{N_i} - l_i} [ N ][ N_i^{\prime} ] 
 \left( \begin{array}{ccc} N_i & 1 & N_i^{\prime} \\ -l_i & 0 & l_i^{\prime} \end{array} \right)
\left( \begin{array}{ccc} N_i & 1 & N_i^{\prime} \\ -M_{N_i} & q_i & M_{N_i}^{\prime} \end{array} \right)
\end{eqnarray}
for $i=1,2$.   The matrix elements of $C_3$ therefore satisfy the selection rules
\begin{eqnarray}
\Delta l = 0, \;\;\;\;| \Delta M_N | \le 1, \;\;\;\; \Delta M_S = 0,
\end{eqnarray}
and so, too, does direct scattering in the Born approximation. It is therefore clear that the dipole interaction will not directly couple the initial state approximated as $ |b, +1; 1, 1/2 \rangle$ to any of the other energetically accessible states listed in the Table.  To make the transition within the Born approximation would require changing the electronic spin.  

However, the electron spin is coupled to the molecular axis, by means of the spin-rotation coupling.  This means that the state nominally labeled $|a, +1; 1, -1/2 \rangle$ in Table \ref{ab_qn} is actually 
perturbed by another states, i.e.
\begin{eqnarray}
|a \rangle \approx |a, + 1; 1,-1/2 \rangle - \frac{ \sqrt{2}\gamma}{d\cal{E}+\gamma} |d, +1; 0,  1/2 \rangle - \frac{2q_l}{d\cal{E}} |e, -1; 1,  -1/2 \rangle,
\end{eqnarray}

Suppose, then, that the initial scattering channel, including partial wave, is
\begin{equation}
\fl
\eqalign{
|{\rm initial} \rangle = |bb \rangle \approx &\Big(\!|b, +1; 1, 1/2 \rangle - \frac{2q_l}{d\cal{E}} |f, -1; 1, 1/2 \rangle\!\Big)\! \Big(\!|b, +1; 1, 1/2 \rangle - \frac{2q_l}{d\cal{E}} |f, -1; 1, 1/2 \rangle\!\Big)\!  \nonumber\\
& \times |L=0, M_L=0 \rangle}
\end{equation}
while the final channel is
\begin{equation}
\fl
\eqalign{
|{\rm final} \rangle = P_{12} |ab \rangle \approx &\, P_{12}  \Big( |a, + 1; -1/2 \rangle - \frac{ \sqrt{2}\gamma}{d\cal{E}+\gamma} |d, +1; 0,  1/2 \rangle - \frac{2q_l}{d\cal{E}} |e, -1; 1,  -1/2 \rangle \Big) \\ \nonumber
& \times \Big(\!|b, +1; 1, 1/2 \rangle - \frac{2q_l}{d\cal{E}} |f, -1; 1, 1/2 \rangle\!\Big) |L=2, M_L = 1 \rangle,
}
\end{equation}
where $P_{12}$ denotes the exchange operator of the two molecules.   Given these states and the selection rules, it is clear that the matrix element $T_{\rm initial, final}$ is nonzero and is proportional to 
${ \gamma}/{(d\mathcal{E}+\gamma)}$,
that is to say, inversely proportional to the electric field.  This final channel is indeed the one that dominates inelastic scattering in the full numerical calculation.  Applying the electric field thus has the effect of reducing the effective spin-rotation coupling of the molecules.  This diminution of effective spin-rotation coupling has been noted previously, in the context of atom-molecule scattering \cite{Tscherbul06_JCP}, and is an important implement in the experimental toolbox for controlling inelastic scattering.



\begin{figure}[h!]
\center
\includegraphics[width=0.8\textwidth]{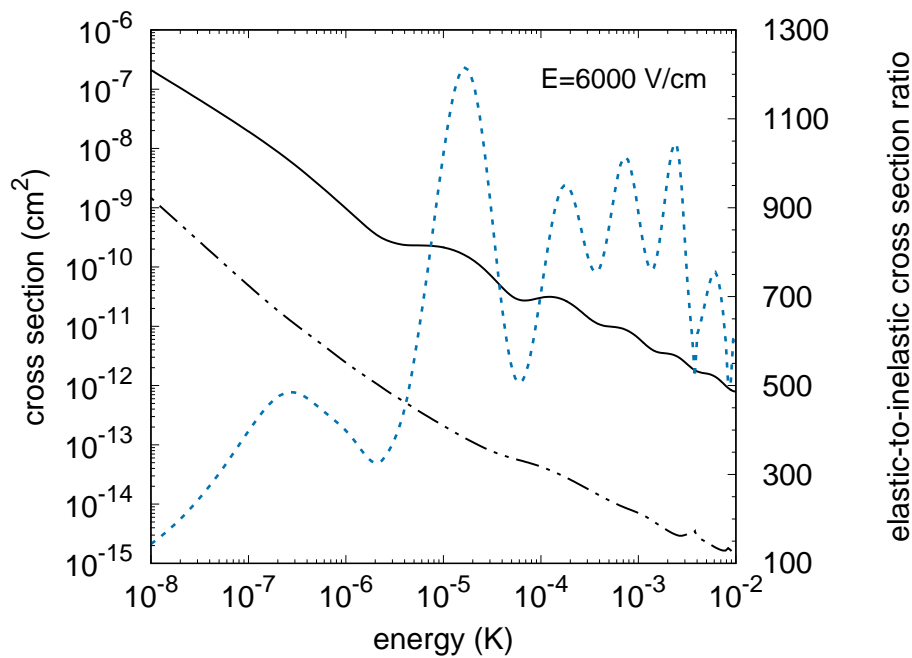}
\caption{Cross sections for elastic (solid curve), inelastic (dash-dotted curve) scattering, and their ratio (dotted curve, right-hand axis) as a function of collision energy. The collision is initiated in the states $|b,l=1,M_F=2\rangle$ of molecules at electric field of $\mathcal{E}=6000$ V/cm.}
\label{crossE6000}
\end{figure}

Because of this suppression, it appears that optically trapped CaOH in the $|b, +1; 1, 1/2 \rangle$ fine structure manifold might be a suitable candidate evaporation. This optimistic message is supported by Figure \ref{crossE6000}, demonstrating elastic $\sigma_{\rm el}$ versus inelastic $\sigma_{\rm inel}$ cross sections as functions of collision energy when electric field is fixed at 6\,000 V/cm. Over the whole energy range, 10\,mK down to 10\,nK, the ratio $\sigma_{\rm el}/\sigma_{\rm inel}$ exceeds 100, even reaching as high as 1000. This ratio becomes smaller towards lower energies because of the Wigner threshold laws that declares $\sigma_{\rm el}$  approaches a constant, while $\sigma_{\rm inel} \propto {E_c}^{-1/2}$ for exothermic collisions.

 \subsection{Field-linked states on collisions of f-state molecules}
 
 The situation is different for molecules initially in the highest fine structure manifold, for example $|f\rangle \approx |f, -1; 1, 1/2 \rangle + 2q_l/d\mathcal{E} |b, 1; 1, 1/2 \rangle$. The state $|f, -1; 1, 1/2 \rangle$ can suffer a direct transition, allowed by the dipole-dipole interaction selection rules, to the energetically lower $|c \rangle$ or $|d \rangle$ state, for large fields well approximated by  
\begin{eqnarray}
\fl
\label{eq:cd}
\Big(|c, -1; 0, 1/2 \rangle + |d, 1; 0, 1/2 \rangle \Big)/\sqrt{2} \quad {\rm{ or }}\quad
\Big(|c, -1; 0, 1/2 \rangle - |d, 1; 0, 1/2 \rangle \Big)/\sqrt{2},
\end{eqnarray}
respectively. Note that in the high-field limit, the states with $M_N=0$ are degenerate between $l= 1$ and $l=-1$, whereby the $l$-doubling Hamiltonian splits them into the linear combinations (\ref{eq:cd}).

In this case, the operator $C_3$ has nonvanishing matrix elements between states $|{\rm initial} \rangle = |ff \rangle $, and $|{\rm final} \rangle = P_{12}|fc \rangle$ or $|{\rm final} \rangle = P_{12}|fd \rangle$, hence the transition is allowed in the Born approximation already. No mixing due to spin-rotation or $l$-doubling is required, and the transition proceeds at a high rate. 

\begin{figure}[h!]
\center
\includegraphics[width=0.8\textwidth]{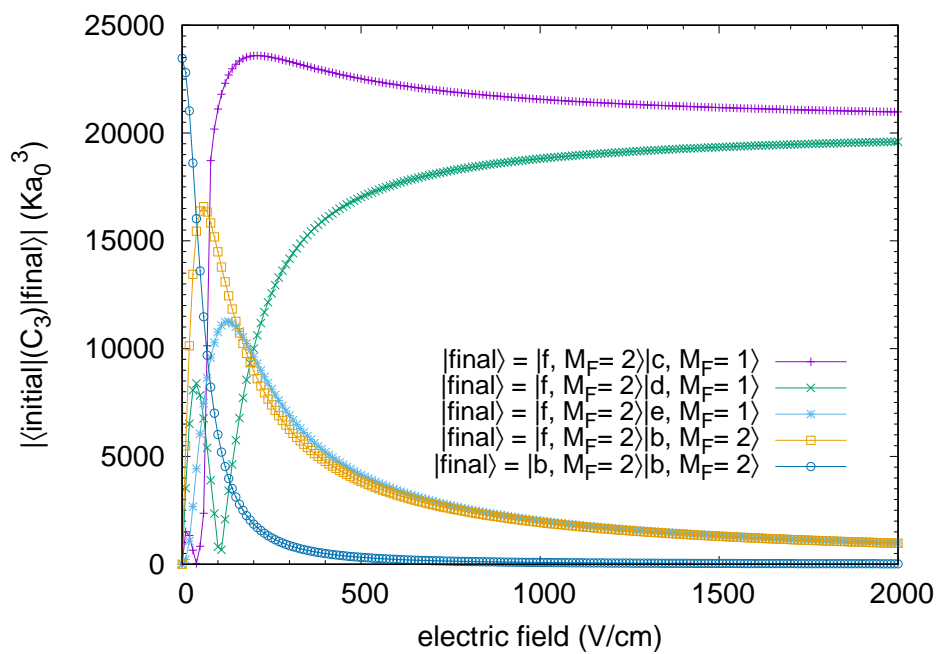}
\caption{Off-diagonal matrix element $|\langle {\rm initial} | C_3| {\rm final } \rangle|^2$ of dipole-dipole interaction without a radial dependence as a function of electric field between the incident channel $|{\rm initial} \rangle =|f,l=-1;M_F=2\rangle|f,l=-1;M_F=2\rangle$, and final channels $|{\rm final} \rangle$ with quantum numbers indicated.}
\label{C3ba}
\end{figure}

For intermediate fields on the order of $\gamma/d$ or larger, states (\ref{eq:cd}) are mixed with other basis vectors $|e, -1; 1, -1/2 \rangle$ and $|a, 1; 1, -1/2\rangle$ due to spin-rotation which allows coupling between states with opposite $M_S$. By contrast, the dipole-dipole operator $C_3$ allows coupling only between states of the same $M_S$, therefore the states $|e, -1; 1, -1/2 \rangle$ and $|a, 1; 1, -1/2\rangle$ will not contribute to $\langle {\rm initial} | C_3| {\rm final } \rangle$ directly, but only via a normalization $[2 + 2\gamma^2/(d\mathcal{E}-\gamma)^2 + 2\gamma^2/(d\mathcal{E}+\gamma)^2 ]^{-1/2}$ of the first-order terms of $|c\rangle$ or $|d\rangle$ states.
Since spin-rotational and $l$-doubling interaction are of similar size, the initial state $|f\rangle$ that involves $|b, 1; 1, 1/2 \rangle$ allows for $C_3$ coupling with the same $l$-manifold state, namely $|d, 1; 0, 1/2 \rangle$.
Hence the dipole-dipole induced transitions behave in an electric field as
\begin{equation}
\eqalign{
\langle ff | C_3 |fc \rangle & \propto \mathcal{N}(\mathcal{E}) \Big(\frac{1}{2} + \frac{|q_l|}{d\mathcal{E}} \Big), \nonumber  \\
\langle ff | C_3 |fd \rangle & \propto \mathcal{N}(\mathcal{E}) \Big(\frac{1}{2} - \frac{|q_l|}{d\mathcal{E}} \Big),
}
\end{equation}
where $\cal{N}(\cal{E})$ is only weakly dependent on $\cal{E}$ due to normalization of initial and final vectors.
Numerically calculated matrix elements of $C_3$ are presented in Figure \ref{C3ba}, in which the hyperfine quantum number notation is restored.  The relevant matrix elements (purple, green) converge down to or up to constant values at large electric field.

Elastic rate coefficients for the $|f\rangle$ states at $E_c=1\,$mK are larger than inelastic rate coefficients but  not by much, certainly not enough for evaporative cooling to occur. For the lower collision energy $E_c=1\,\mu$K the inelastic scattering rates increase rapidly at low electric field then level off at about 100 V/cm.  This behavior attests to the induction of dipole moments by the field, which then increase the ability of the  molecules to exert torques on one another and change their internal state.  

 \begin{figure}[h!]
 \center
\includegraphics[width=0.8\textwidth]{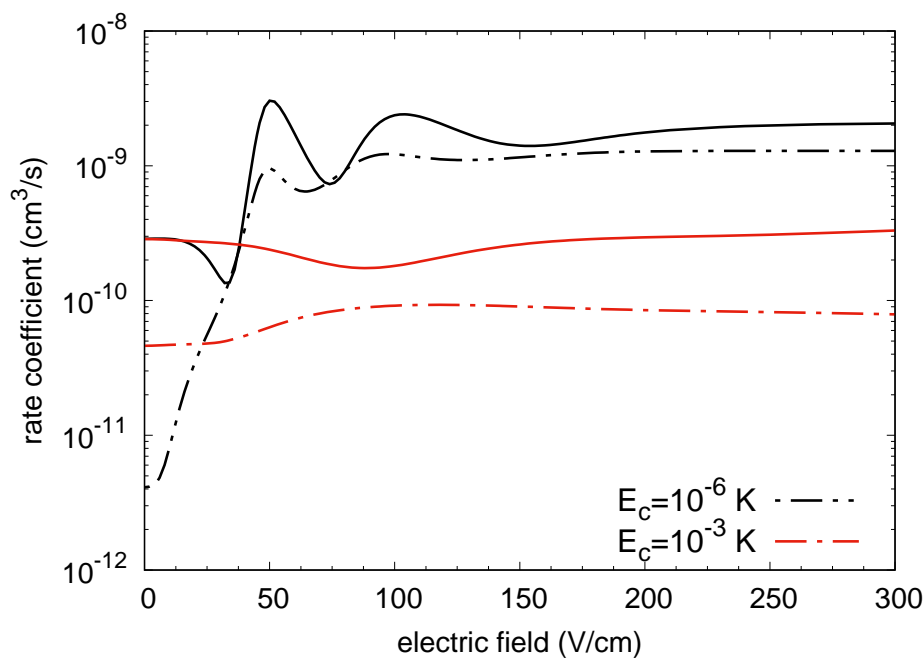}
\caption{Rate coefficients for elastic (solid curves) and inelastic (dashed curves) scattering as a function of electric field. The collision is initiated in the states $|f,l=-1,M_F=2\rangle$ of molecules at two different collision energies $E_c = 1 \mu$K (black lines) and $E_c = 1$ mK (red lines).}
\label{rateE10-36_f}
\end{figure} 

In addition, the collision rates exhibit modulations as the field is turned on, which are more pronounced at the lower energy.  These modulations correspond to a set of ``field-linked'' resonant states, anticipated in scattering of dipolar $^2\Pi$ molecules \cite{Avdeenkov03_PRL}.  They correspond to long-range, quasi-bound states of the two molecules.  The resulting (CaOH)$_2$ dimer is held by a delicate balance between the attractive and repulsive aspects of the dipole-dipole interaction, and exist only in the presence of an electric field that activates these dipoles; hence the name field-linked.  Details on the structure of these dimers are described in \cite{Avdeenkov04_PRA}.

\begin{figure}[h!]
\center
(a)\includegraphics[width=0.6\textwidth]{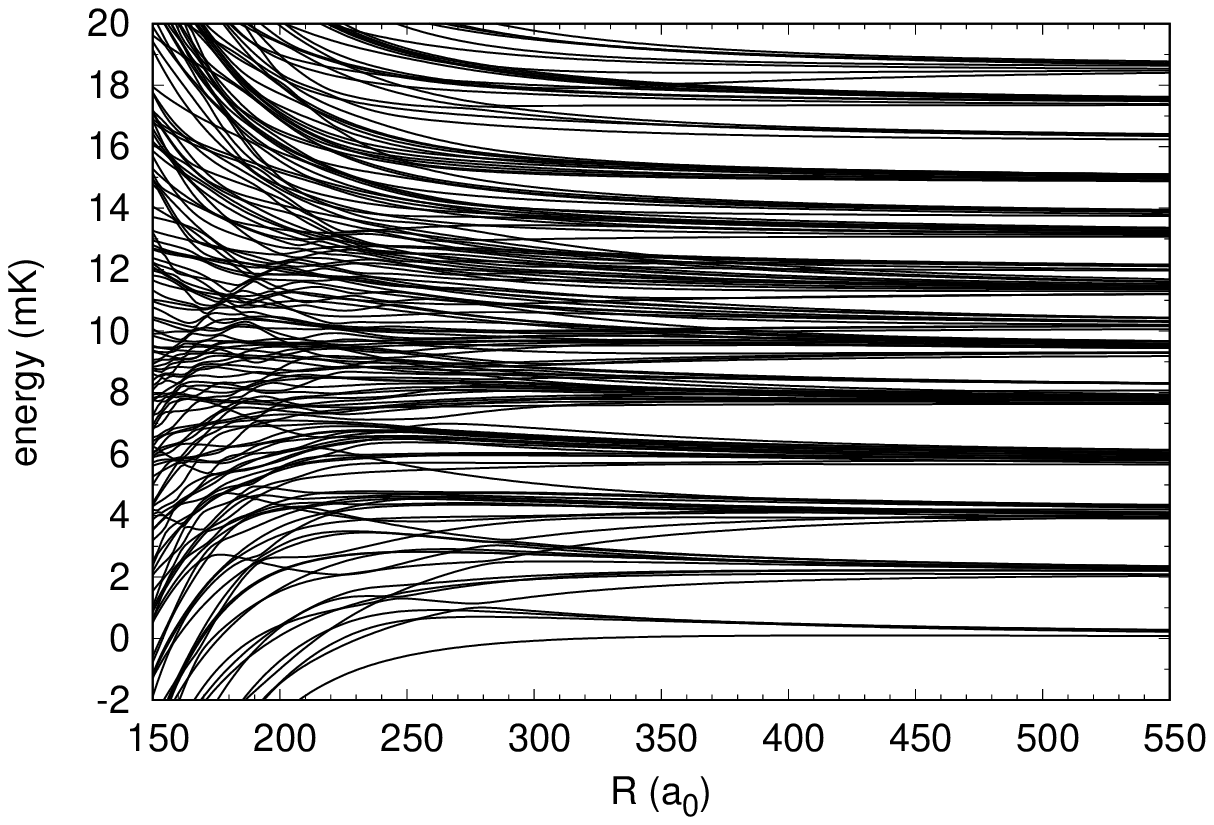}
(b)\includegraphics[width=0.6\textwidth]{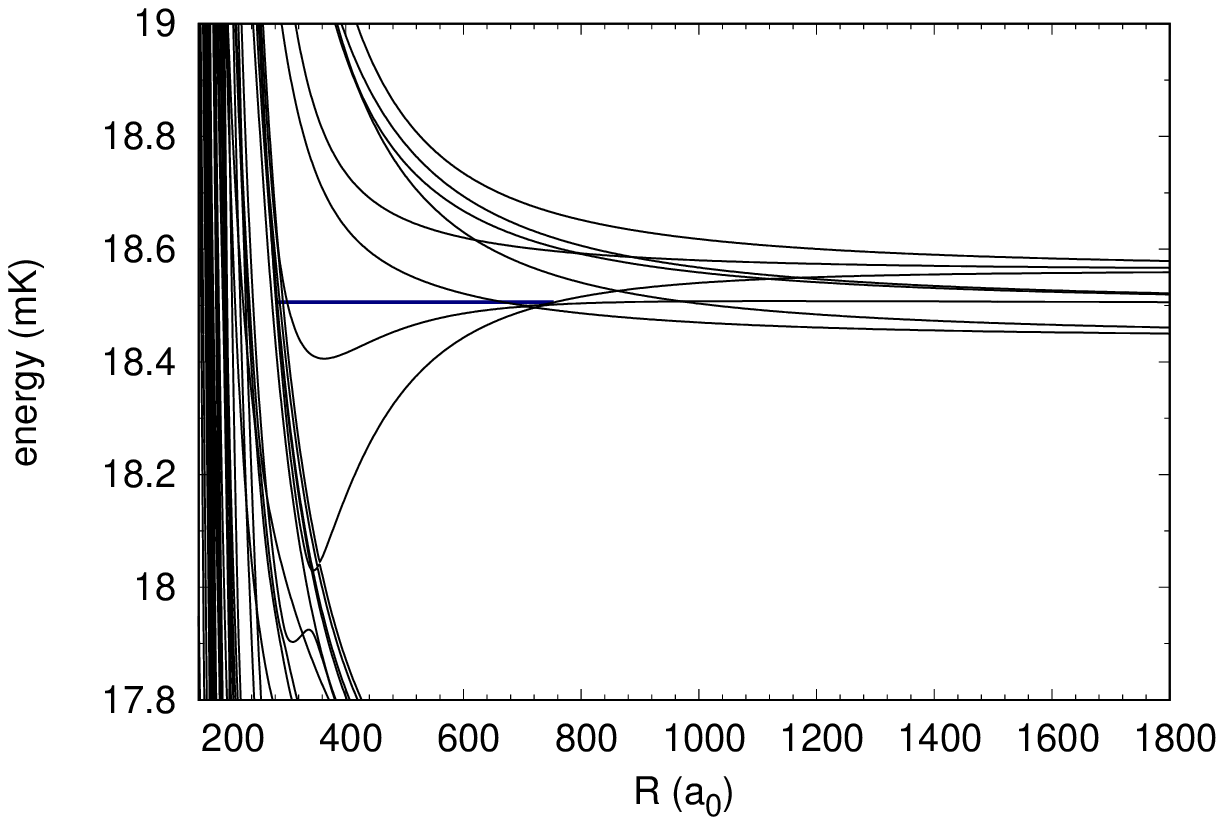}
\caption{(a) Adiabatic curves of potentials for $L=0,2,4$ at fixed values of electric field 195 V/cm.  Panel (b) is a zoom of panel (a) for energies that show $ff$ thresholds, the most upper channel cluster of panel (a). Blue heavy line corresponds to energy of a quasi-bound state (see text).}
\label{Adiabcurves}
\end{figure}

These resonant states represent an oasis of relative simplicity amid the chaos of ultracold molecule interactions.  Figure \ref{Adiabcurves}a) shows a partial set of the adiabatic curves at an electric field value ${\cal E}=195$ V/cm, near the peak of the modulation of $K_{\rm el}$ in Figure \ref{rateE10-36_f}.  For clarity, only those channels dominated by the partial waves $L=0,2,4$ are shown.  A great deal of fine- and hyperfine structure appears, along with many multiple crossings.  However, in the vicinity of the very highest threshold, correlating to pairs of molecules in the $|f, -1; 1, 1/2 \rangle$ state, one sees a potential energy curve with a minimum at around 340 $a_0$ and an inner turning point near 280 $a_0$ [Figure \ref{Adiabcurves}b)]. This potential cradles the filed linked states, which are relatively isolated from the rest of the spectrum.  

Field-linked (CaOH)$_2$ dimers could presumably be produced, by ramping the electric field from low to high values across the resonance, adiabatically converting molecules to dimers in the same way that alkali atom are converted to Feshbach molecules upon sweeping a magnetic field.  The dimers could then serve as a platform for further manipulation of molecular interactions, for example, selective laser excitation that could probe the reaction barrier, or else Raman processes that could create selected states of the dimer.  A key feature of the field-linked dimer is that its lifetime is short, since the polarized molecules continue to exert torques on one another.  

To determine this lifetime, we compute the Wigner-Smith time delay \cite{Smith60_PR}.  We begin by computing the energy-dependent eigenvalues $K_i(E)$ of the scattering $K$-matrix to obtain the eigenphase shifts $ \delta_i (E)= \tan^{-1}K_i(E)$.  The eigenphase sum,
\begin{eqnarray}
\delta(E) = \sum_i \delta_i(E)
\end{eqnarray}
is a quantity that rises by $\sim \pi$ as the energy crosses a resonance.  The sharper this rise,  the narrower the resonance, and the longer the lifetime.  Formally, the time delay is given by
\begin{eqnarray}
\tau = 2 \hbar \frac{ d \delta }{ dE }.
\end{eqnarray}
The time delay peaks at resonant energies, and its value at the peak is associated with the lifetime of the resonance.  

\begin{figure}[h!]
 \center
(a)\includegraphics[width=0.6\textwidth]{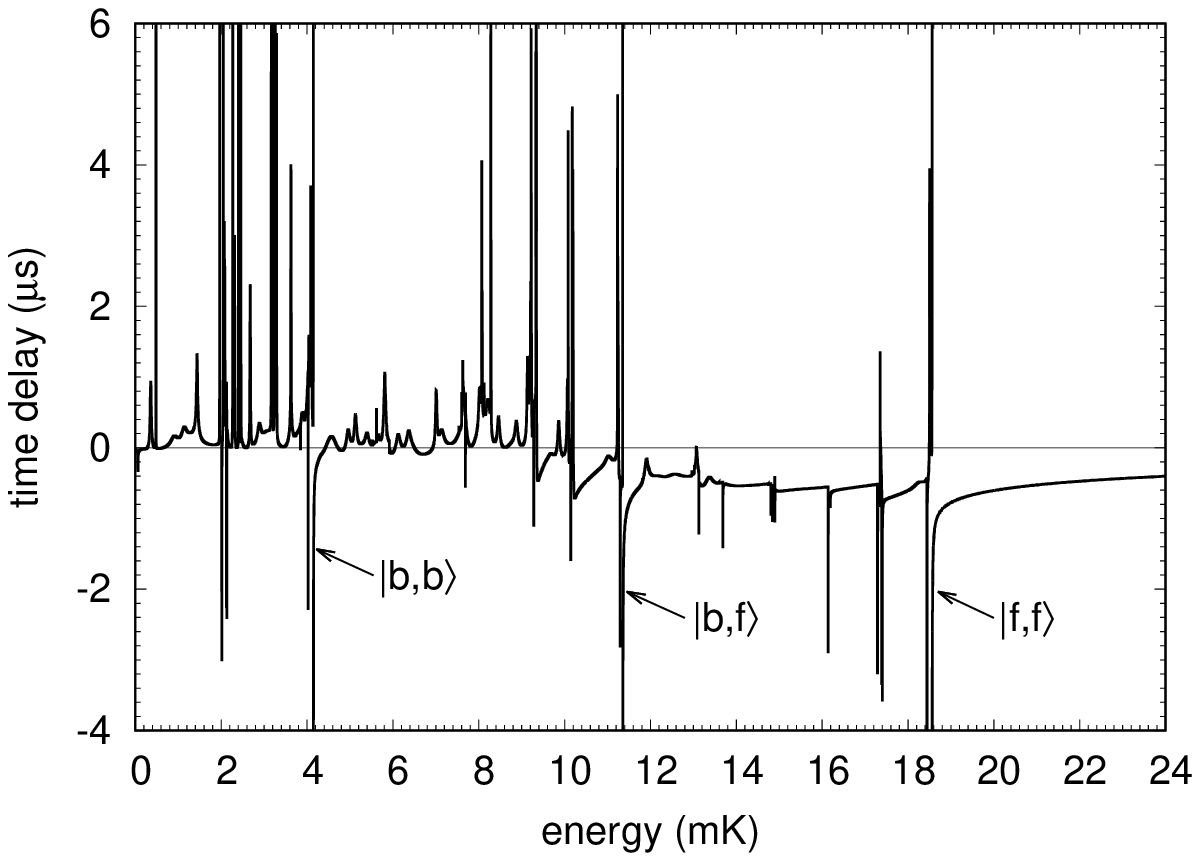}
(b)\includegraphics[width=0.6\textwidth]{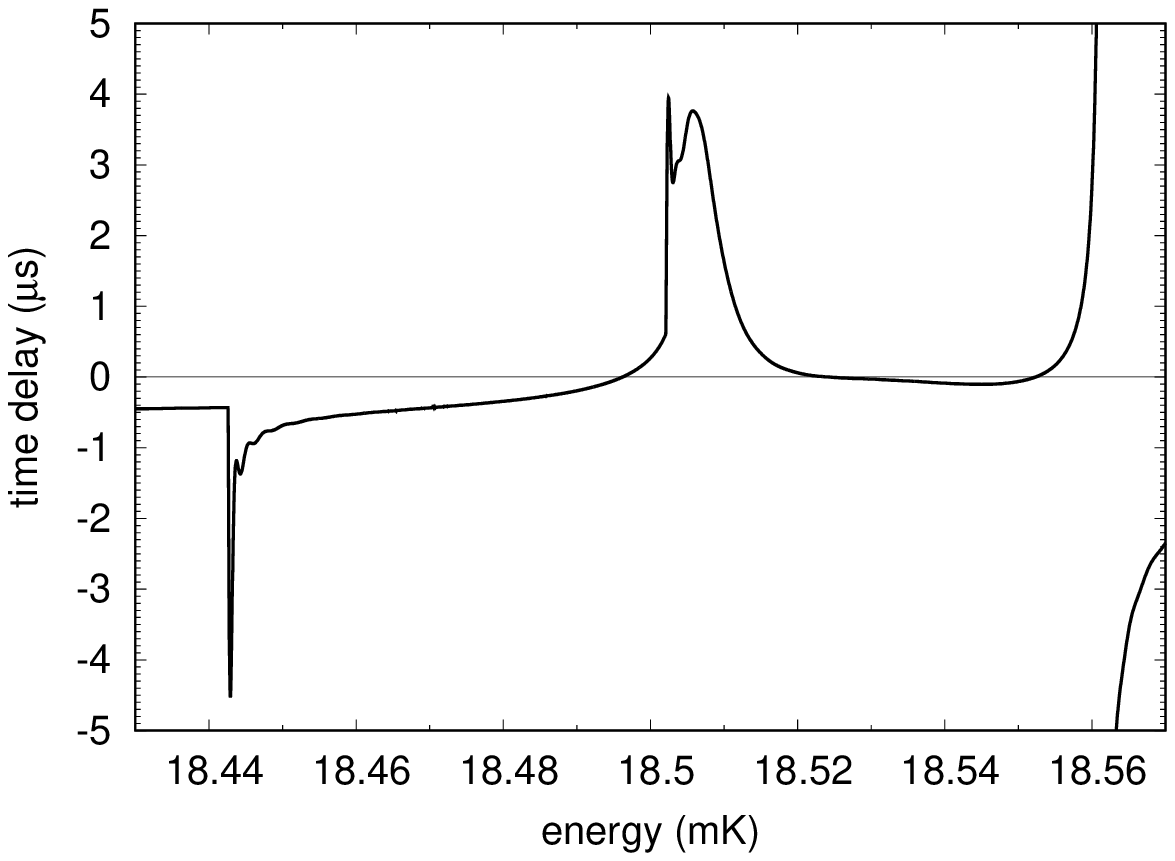}
\caption{(a) Time delay versus energy for scattering of molecules in their stretched $|f, l=-1; M_F = 2\rangle$ state at $\mathcal{E}=195$ V/cm. Number of partial waves is here reduced to $L=0,2,4$. Threshold energies of interest are labeled. (b) Same as in (a) but in detailed resolution to reveal the field-linked resonant state.}
\label{timedelay}
\end{figure}

Figure \ref{timedelay}a) presents $\tau$ at ${\cal E}=195$ V/cm over an energy range up to 24 mK. For reference, three fine-structure thresholds are shown.  At low energies, many resonances are seen.  These are primarily Fano-Feshbach resonances with the many hyperfine states.  The resonant wave functions in this energy range penetrate to small $R$, given the many attractive adiabatic curves in this range [Figure \ref{Adiabcurves}a)], and are therefore poorly characterized by the model. For energies larger than about 10 mK, time delay is generally negative because the particle spends less time in the short range as it is reflected from the repulsive  potential energy curves  shown in Figure \ref{Adiabcurves}a). 

 At the energy $\sim$18.506 mK, just below the $|f, -1; 1, 1/2 \rangle  |f, -1 ;1, 1/2 \rangle$ threshold of interest, the time delay exhibits a striking resonance peak isolated from other resonances. This is the signature of the field-linked state, and its peak time delay is $\sim 4$ $\mu$s.  We conclude that, upon formation, the (CaOH)$_2$ field-linked dimer would live for several microseconds, giving the experimenter time to further manipulate the molecules.

 \section{Conclusions}
 
 Various aspects of the long-range physics between dipolar molecules, predicted but never observed for dimers like OH, are shown to occur also in the $(0,1^1,0)$ $N=1$ states of CaOH.  The big difference is that previously considered molecules have been produced by buffer gas cooling, Stark deceleration, or other methods that limited their ultimate temperature to the 10-100 mK regime.  By contrast, the novel ability to laser cool species such as CaOH opens the possibility that these intricate effects can be measured and exploited to further the development of ultracold molecular science.  
 
 We have focused on two of the main features of these ultracold collisions.  One the one hand, by choosing spin-stretched $b$ state molecules, the elusive goals of evaporative cooling and even dipolar molecular quantum degenerate gases may be achieved.  In addition, by choosing $f$ state molecules, novel field-linked dimers become possible, opening new implications for studying and  manipulating the molecules.  
 
\ack
This material is based upon work supported by the
National Science Foundation under Grant Number PHY
1125844 and Grant Number PHY 1806971. L.D.A. acknowledges the financial support of the Czech Science Foundation (Grant No. 18-00918S).

\section*{References}
\bibliographystyle{iopart-num}
\bibliography{Bibliography}

\end{document}